      [1999/12/01 v1.4c Il Nuovo Cimento]
\documentclass{cimento}

\usepackage{epsfig}
\usepackage{bm}

\title{Top forward-backward asymmetry 
in chiral $\bm{U(1)^\prime}$ models}
\author{Pyungwon~Ko\from{ins:KIAS},
Yuji~Omura\from{ins:KIAS}
        \atque
Chaehyun~Yu\from{ins:KIAS}\thanks{Speaker.}}
\instlist{\inst{ins:KIAS} School of Physics, KIAS, Seoul 130-722, Korea
                          }
\PACSes{\PACSit{12.60.-i, 14.65.Ha, 14.80.-j}
{Models beyond the standard model, Top quarks, 
 Other particles (including hypothetical).}
}
\begin{document}

\maketitle

\begin{abstract}
We construct flavor-dependent chiral $U(1)^\prime$ models 
with a $Z^\prime$ boson which couples to the right-handed up-type quarks
in the standard model (SM). To make the models have realistic renormalizable
Yukawa couplings, we introduce new Higgs doublets with nonzero $U(1)^\prime$ 
charges. Anomaly-free condition can be satisfied by adding extra chiral
fermions. We show that these models could analyze the discrepancy
between the SM prediction and empirical data in the top forward-backward 
asymmetry at the Tevatron. 

\end{abstract}

During the last three years, the discrepancy between the standard model (SM)
prediction and empirical data in the top forward-backward 
asymmetry ($A_{FB}$) at the Tevatron has drawn much attention in particle
physics. The SM QCD prediction is $0.072^{+0.011}_{-0.007}$ 
at NLO+NNLL accuracies~\cite{NNLL}.
While the combined CDF and D0 results are $0.210\pm 0.067$  and
$0.196\pm 0.060^{+0.018}_{-0.026}$, respectively, both of which are deviated
from the SM prediction by about $2\sigma$~\cite{AFB}.
Among a lot of theoretical attempts to resolve the discrepancy~\cite{models}, 
the light leptophobic $Z^\prime$ model with large flavor changing couplings 
to the right-handed (RH) up-type quarks has been paid much attention 
to~\cite{zprime} because of its large flavor changing
neutral currents. This rather phenomenological model is strongly
disfavored by the same sign top pair production constraint 
at the LHC~\cite{same}. However we point out 
that this simple $Z^\prime$ model is not a realistic model. 
The $Z^\prime$ boson should be associated with, for example,  
a chiral $U(1)^\prime$ symmetry~\cite{u1}, whose charge
assignment is flavor-dependent. Then, in order to write down proper Yukawa 
interactions, additional Higgs doublets with nonzero $U(1)^\prime$ charges
must be included~\cite{chiralu1}. 
This is inevitable when the new gaue group is chiral.
In general such a flavor-dependent chiral $U(1)$ symmetry yields gauge anomaly, 
which can be canceled by introducing
extra fermions~\cite{chiralu1}. After breaking both $U(1)^\prime$ and
EW symmetries, the $Z^\prime$ and Higgs bosons could contribute to 
the $t\bar{t}$ pair production with large $A_{FB}$. Interference between
the $Z^\prime$ and Higgs bosons can relieve the strong constraint
from the same sign top pair production. We emphasize that the realistic
models with a leptophobic $Z^\prime$ boson which couples 
to the RH up-type quarks accompany modification of the Higgs sector
and its effects must be taken into account together with the $Z^\prime$ boson, 
in particular, at the LHC. This is true not only in the $Z^\prime$ model,
but also in the other models~\cite{AFB} designed to account for $A_{FB}$ 
at the Tevatron with flavor-dependent couplings.


We assume that only the RH up-type quarks ($U_R^i$, $i=1,2,3$) are charged 
under $U(1)^\prime$ with its charge $u_i$ while the left-handed quarks 
($U_L^i$ and $D_L^i$) and the RH down-type quarks ($D_R^i$) are uncharged.
In the mass eigenstates $\hat{U}_R^i$, the interaction Lagrangian for the
$Z^\prime$ and $\hat{U}_R^i$ is given by
$
\mathcal{L} = g^\prime Z^{\prime \mu}
\left\{ (g_R^u)_{ij} \overline{\hat{U}_R^i} \gamma_\mu \hat{U}_R^j 
\right\},
$
where $g^\prime$ is a gauge coupling of $U(1)^\prime$
and $(g_R^u)_{ij} = (R^u)_{ij} u_k (R^u)_{kj}^\dagger$ with 
the biunitary matrix $(R^u)_{ij}$ diagonalizing the up-type quark mass 
matrix~\cite{chiralu1}. In general $(g_R^u)_{ij}$ could have nonzero 
off-diagonal elements.

As we have discussed, we have to introduce new Higgs doublets $H_i$
with $U(1)^\prime$ charges $-u_i$ in order to get realistic renormalizable 
Yukawa couplings. The number of Higgs doublets including the SM Higgs
depends on the charge assignment. One can choose the two-Higgs doublet case 
with $u_i=(0,0,1)$, the three-Higgs doublet case 
with $u_i=(-1,0,1)$, or \rm{etc}.
In the mass basis, the Yukawa interactions of lightest neutral scalar Higgs $h$
and pseudoscalar Higgs $a$ can be given by
\begin{equation}
\mathcal{L} = - Y_{ij}^u \overline{\hat{U}_{Li}} \hat{U}_{Rj}h
+ i Y_{ij}^a \overline{\hat{U}_{Li}} \hat{U}_{Rj} a + h.c.,
\end{equation}
where the Yukawa couplings $Y_{ij}$ and $Y_{ij}^a$ are functions 
of $g_R$, $m_i^u$ and the mixing angles and vacuum expectation values 
of the neutral Higgs fields and can have off-diagonal elements~\cite{chiralu1}.
Finally the gauge anomaly can be canceled by introducing an extra generation
and two pairs of $SU(2)$ doublets of $SU(3)$ triplets 
as in Ref.~\cite{chiralu1}.


\begin{figure}
\includegraphics{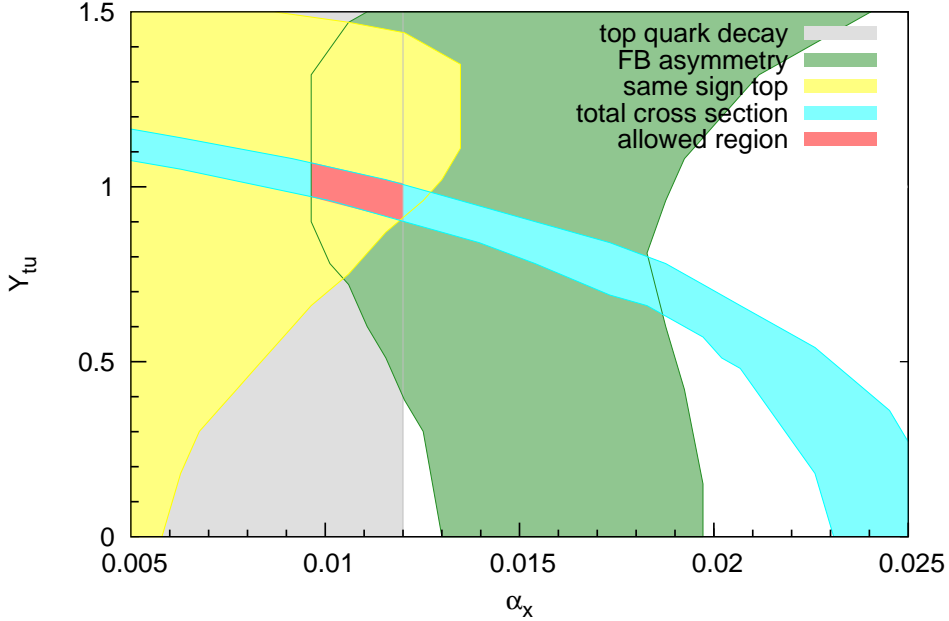}     
\caption{
The favored region for $\alpha_x$ and $Y_{tu}$.
}
\label{fig1}
\end{figure}

In our model, it could be expected to have large off-diagonal Yukawa couplings
$Y_{tu}^{(a)}$ because they are proportional to the top quark mass.
The mixing matrix $(g_R^u)_{ij}$ could generally have nonzero off-diagonal
elements. We assume that $(g_R)^u_{13}\equiv (g_R^u)_{ut}$ is large, but
the other components are well suppressed. Then both the $Z^\prime$ boson and 
Higgs bosons $h$ and $a$ can contribute to the $t\bar{t}$ production
through the $u\bar{u}\to t\bar{t}$ process.
The $t\bar{t}$ production rate at the Tevatron and LHC is in good
agreement with the SM. We impose that the $t\bar{t}$ cross section 
at the Tevatron in our model with a $K$ factor $K=1.3$ 
is consistent with the measurement
$(7.5\pm 0.48)$ pb~\cite{cdf-note-9913}. In the SM the top quark 
dominantly decays to the $Wb$. It might be dangerous that the branching
fraction of the top quark decay to the non-$Wb$ state, which can be generated
by the flavor changing neutral currents, is too large. Thus we restrict 
the branching fraction of the non-SM decay of the top quark to 5\%.
The flavor changing couplings in the top sector bring on the same sign
top pair production through the $u u \to t\bar{t}$ process
and the single top production through the $g u \to t+X (X=Z^\prime,h,a)$ 
process. We require the bound $\sigma(pp\to tt) < 17$ pb from the CMS
experiment~\cite{same}, but the single top production data give 
no constraints to our model because of difference between final states
in experiments~\cite{singletop} and our model. 
Lastly we consider $A_{FB}=(0.158\pm0.075)$
in the lepton+jets channel at CDF~\cite{AFB}. As an illustration of our model
we take $m_{Z^\prime}=145$ GeV $m_h=180$ GeV, $m_a=300$ GeV, 
and $Y_{tu}^a=1.1$. 

In fig.~\ref{fig1}, we show the allowed region
for $\alpha_x\equiv (g^\prime {g_{R}}_{ut})^2/(4\pi)$ and $Y_{tu}$.
The gray, cyan, green, and yellow regions correspond to the regions allowed
by constraints from the top quark decay, 
the $t\bar{t}$ production cross section
at the Tevatron, $A_{FB}$ at CDF, and the same sign top pair production
at CMS, respectively. The red region is favored from the empirical data
at the Tevatron and LHC. It is remarkable that the strong constraint 
from the same sign top pair production can be relieved in our model 
due to the destructive interference 
between the $Z^\prime$ boson and Higgs bosons.

\begin{figure}
\includegraphics{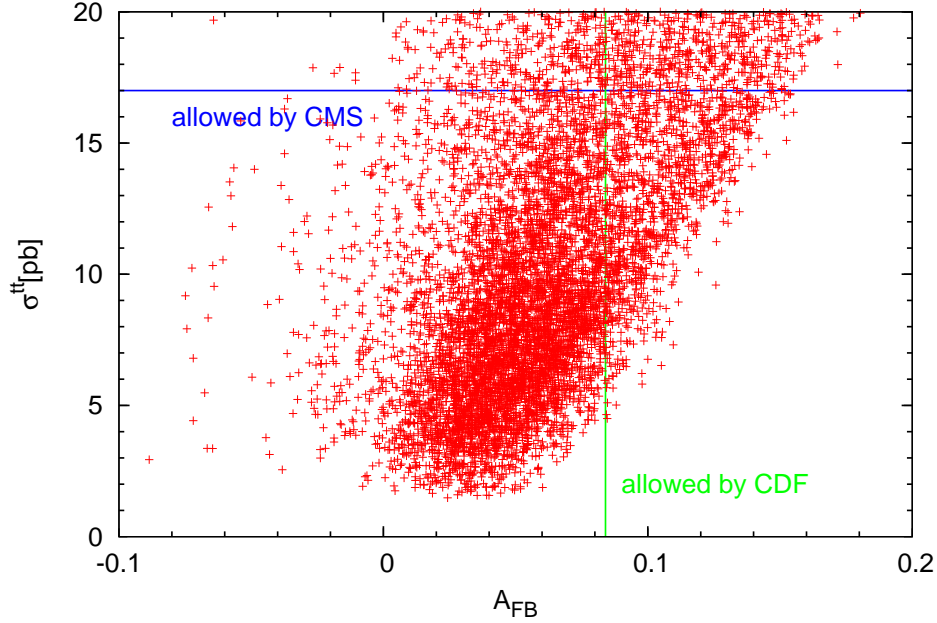}     
\caption{
The scattered plot for $A_\textrm{FB}$ at the Tevatron and
$\sigma_{tt}$ at the LHC in unit of pb.
}
\label{fig2}
\end{figure}

Now we examine our model by varying the model parameters. In this analysis
we fix $m_{Z^\prime}=145$ GeV, but we choose the following parameter regions:
$180~\textrm{GeV} < m_h, m_a < 1~\textrm{TeV}$, 
$0.005 < \alpha_x < 0.025$, and
$0.5 < Y_{tu}, Y_{tu}^a < 1.5$ with the constraint $|Y_{tu}|<|Y_{tu}^a|$.
In fig.~\ref{fig2}, we show the scattered plot for $A_{FB}$ at the Tevatron and 
the cross section $\sigma^{tt}$ for the same sign top pair production 
at the LHC. All the red points in fig.~\ref{fig2} satisfy the $t\bar{t}$
cross section rate at the Tevatron. In fig.~\ref{fig2}, 
the points in the right-lower side are allowed by the empirical data.
The points in the allowed region have the following bounds:
$180~\textrm{GeV} < m_h < 250~\textrm{GeV}$, 
$0.005 < \alpha_x < 0.014$, 
$0.75 < Y_{tu} < 1.3$, and $0.9 < Y_{tu}^a < 1.5$. $m_a$ is not constrained by
the data. We note that about 7\% $A_{FB}$ from the SM NLO contribution
is ignored in this work and it could further enhance $A_{FB}$ at the Tevatron.

In conclusions, we presented $U(1)^\prime$ flavor models with
flavor dependent $Z^\prime$ couplings only to the RH up-type quarks.
Additional Higgs doublets with $U(1)^\prime$ charges
were introduced in order to to construct realistic models
to have renormalizable Yukawa couplings for the up-type quarks. 
For cancellation of the gauge anomaly we included the extra fermions 
charged under $U(1)^\prime$.
We showed that our model can explain the forward-backward asymmetry
measured at the Tevatron and evade the stringent constraint 
from the same sign top pair production at the LHC through the destructive
interference between the $Z^\prime$ boson and Higgs bosons.
We anticipate that our model will be probed in the future experiments.


\acknowledgments
This work is supported by Basic Science Research Program
through the National Research Foundation of Korea (NRF) funded by the
Ministry of Education Science and Technology (2011-0022996).


\end{document}